# Topological Structures of Cluster spins for Ising Models

You-Gang Feng

Department of Basic Sciences, College of Science, Guizhou University, Cai-jia Guan,

Guiyang, 550003 China

E-mail: ygfeng45@yahoo.com.cn

   Widom was first pointed out that, as the distance from the critical point is varied, thermodynamic functions change their scale but not their function forms, an idea of scaling was proposed. Kadanoff applied the idea to the Ising model and in so doing opened the way for the modern theory of critical phenomena introduced by Wilson. A concept of cluster spin was introduced by them. A cluster spin is an ordered block containing lattices and keeping the symmetries of the original system. The method introduced a concept of self similar transformations. The renormalization group theory indicates that the critical point corresponds to the fixed point of the self similar transformations. In their work it is necessary to calculate interactions of lattice spins in a cluster. The greater a cluster size, the more great and complex the work will be, which forced authors to select some approximate methods. Two rules are often used, the coarse graining and the decimation, which character is to obtain a cluster spin by making use of non-product of lattice spins in the cluster. These approximate methods resulted in that the authors never gave us any highly accurate solutions. Moreover, in the mathematics sense, the rules bring us a mathematical paradox: On the one hand, the non-product itself implies that each lattice spin in the cluster is independent, there is an empty set between two nearest-neighbor lattices, so each lattice spin is a connected component and the cluster can be regarded as a multi-tuply connected space rather than a simply connected one, thus the space cannot shrink to a lattice (spin) with the help of the non-product of its components. On the other hand, the cluster is made to be a spin as a result of these rules, which indicated that the cluster is simply connected. We think that for a system its physical laws relate tightly to its topological properties, if we violate its topological patterns we will not obtain real physical characters.

The renormalization group theory indicates that there are self similar transformations in Ising model at the critical temperature, which fixed point corresponds to the critical point. Some authors supposed that only if the size of a cluster becomes infinite the exact solution can be obtained. Unfortunately, it is not true. The reason is explained mathematically as follows. Let the clusters sizes and shapes to be identical, so the system can make the self similar transformations entirely. Let the distance of interaction between two nearest neighbor cluster spins $A$ and $B$ be $d(A,B)$, after a self similar transformation they become nearest neighbor lattice spins $f(A)$ and $f(B)$, and the distance of interaction between neighbors be $d(f(A), f(B))$, the Lifschetz constant $L$ be defined by

$$L = \frac{d(f(A), f(B))}{d(A,B)} \tag{1}$$

If the cluster size is limited, i.e. $d(A,B) < \infty$, so that $0 < L < 1$ for $d(A,B) > d(f(A), f(B))$. If the cluster size approaches infinity, so $d(A,B) \to +\infty$ and the constant tends to zero, $L \to 0$. According to Lifschetz fixed point theorem, there is not any fixed point in the self similar transformations for $L \to 0$. If and only if the cluster size is limited the constant $L$ will certainly be smaller than one and not vanish, $0 < L < 1$, so that the self similar transformations of the cluster spins have a unique fixed point. At that time the correlation length in a system changes into infinity only by infinite iterations, which shows that the self similar transformations are under the necessity of hierarchies, and on the infinite hierarchy the system will become an ordered cluster. We call an original lattice the zeroth order lattice, they construct a first order cluster by the transformation, and the cluster is said to be on the first hierarchy of the transformations. On the $m$ th hierarchy there must be only the $m$ th order clusters which are independent of each other, and a $m$ th order cluster contains the ($m$-1)th order lattices which are just the ($m$-1)th order clusters before rescaling. It is clear that the inside space of the $m$th order cluster is just the outside space of the ($m$-1)th lattices, whish is of dimensions $D$. As a lattice, the inside of a ($m$-1)th order lattice is indistinguishable due to the rescaling rule.

There are varieties of lattice systems, their clusters differ from each other, what characters the clusters will have? A carrier space of an ordered cluster must be simply connected tantamount to a single point space, because of which an ordered cluster can shrink to a lattice. This character is a mathematic basis for the rescaling cluster spins. Topology point out that only triangles and tetrahedroids are simply connected, they are mathematically called simplexes, without relating to their sizes. By connectivity we divide lattice systems into two classes: The first, an irreducible system to which only the plane triangle lattice and the tetrahedroid lattice belong. The second, a reducible system, to which all of lattice systems belong apart from the first. In the irreducible system a cluster which keeps the system symmetries is called irreducible cluster, the triangle clusters and the tetrahedroid clusters are of them. Some identical irreducible clusters can directly form a new larger irreducible cluster, making the transformations go on. In the reducible system a cluster which keeps the system symmetries is called reducible cluster, of which a square lattice and a cube lattice are. Although a reducible cluster keeps all of the symmetries of the original system, but its carrier space is a complex in the mathematic sense, which is not simply connected. Mathematically, a complex can be decomposed to some simplexes, each of which is simply connected and is called a subcluster of the reducible cluster. The subclusters shape must differ from the reducible cluster's one, otherwise they are merely the reducible clusters with different sizes and the same topological properties. The subclusters keep only partial symmetries of the original system, if they execute directly the transformation the original symmetries will be broken, thus such a transformation is impossible. The self similar transformations of a reducible system take two steps: First, $k$ subclusters formed in a reducible cluster. Second, the $k$ subclusters make the reducible cluster ordered by their interactions. In fact, the two steps occur simultaneously. After that, some ordered reducible clusters can construct a new bigger reducible cluster, making the self similar transformations on a higher hierarchy. Topology indicates that a carrier space of an ordered reducible cluster is equivalent to a product space of its subclustres' spaces. A geometric grid is a carrier of lattice spins and the topological properties of the spin system are attached to the topo-logical ones of the system's carrier. See figure 1, a triangle with vertices of $P$ has $n^2$ cells, $P = (n+1)(n+2)/2$, where a cell is a minimal simplex. If we put a spin on each vertex, the simplex then becomes a cluster spin containing $P$ lattice spins with edge $n$, where a distance of two nearest neighbors is defined as a unit length.

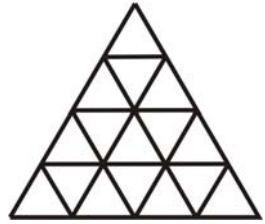

Fig. 1.  An irreducible cluster of the triangle lattice with edge $n = 4$.

The self similar transformations have fractal properties, which leads to fractal dimensions. Wilson did not notice that a cluster is of fractal structure, the difficulties in calculations forced him to take some approximate methods. We think that understanding the fractals will help us research deeply the properties of the critical points. In the triangle lattice system the spin directions on the lattice sites differ from each other, such a non-uniformity of the orientations makes a lattice different from others, thus a cluster has a fractal dimension. Let a cluster with edge $n$ be covered by open balls with diameter $1/n$, which number be $P$ at least, then the fractal dimension $D$ of the cluster is defined as

$$D = -\frac{LnP}{Ln(1/n)} = \frac{LnP}{Ln(n)} \qquad (2)$$

Both a cluster and a fractal dimension are produced in the self similar transformation, as mentioned previously the cluster is an ordered block, so only the cluster has the fractal dimension. A disordered block acts as an empty set in the transformation process, and there is no relation between a disordered block and a fractal dimension. The values of the edges should guarantee the fractal dimension meaningful, otherwise (2) has no meaning. By the definition (2), the inside space of a cluster amounts to a super cube of dimensions $D$ with edge $n$ and volume $P = n^D$, in this case the fractal dimensions are also called capacity dimensions. By (2), a fractal dimension $D_{tr}$ for the irreducible cluster of the triangle lattice is

$$D_{tr} = \frac{Ln[(n+1)(n+2)/2]}{Ln(n)} \tag{3}$$

From (3) we see that $D_{tr}$ is an edge function. For the plane square lattice, since there is no next-nearest neighbor interaction in Ising model the carrier of a reducible cluster cannot be decomposed to some triangles, as usual. However, as the carrier is complex, it will be decomposed certainly to some special shapes. Let there be $k$ subclusters in a reducible cluster, from (2) a fractal dimension $D_{sq}$ of a subcluster of the square lattice be defined by

$$D_{sq} = \frac{Ln[(n+1)^2/2]}{Ln(n)} \ , \tag{4}$$

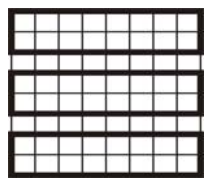

Fig.2. A reducible cluster containing three subclusters of the square lattice with $n = 8$.

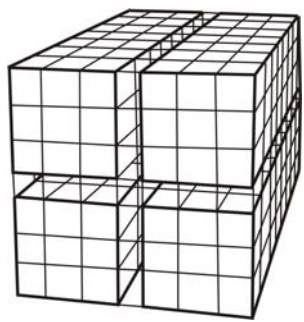

Fig. 3. A reducible cluster with four subclusters of the cube lattice with $n = 7$.

where $k=2$. In figure 2, $k=3$, the subcluster intervening between two subclusters cannot transform in the same way as the other subclusters do. When the edges of its neighbor subclusters change into half-infinity, the edge of the middle subcluster has to keep limited, such non-uniform transformation will break the original symmetries. If $k=4$, there are four squares with smaller edges which are complexes equally. The cases of $k>4$ are similar to the cases of $k=3$ or $k=4$. Similarly, the fractal dimension $D_{cu}$ of a subcluster of the cube lattice is

$$D_{cu} = \frac{Ln[(n+1)^3/4]}{Ln(n)} \qquad (5)$$

As illustrated in figure 3, a cube is subdivided into four cuboids, $k=4$. According to simplicial decomposition theory a three dimensional complex should be composed of three dimensional simplexes, namely, it must be decomposed in three dimensional directions, so $k \neq 2$. If $k=8$ there are 8 cubes with smaller edges which are complexes still. If $k=3$, 5, 6, 7 or $k>8$ the transformations will break the system symmetries. Thus the case of $k=4$ is correct. The changing of the fractal dimension reflects directly the self similar transformations. Since the fractal dimension is an edge function the edges fixed point certainly accords with the transformations' one. Now that different sizes accord with different fractal dimensions, for those clusters with different fractal dimensions what values of spins they will have? Let a cluster spin be $S$, the energy of interaction between two nearest neighbors denoted by $y_1$ and $y_2$ be $JS^2$, where $J$ be a coupling constant; as the self similar transformation a new cluster formed on higher hierarchy the two cluster spins become two nearest neighbor lattice spins in the new cluster after rescaling, and denoted by $f(y_1)$ and $f(y_2)$ with each magnitude of $s$, $s^2=1$, and their interacting energy be $js^2$, where $j$ be a coupling constant in the $D$-dimensional space, the inside space of the new cluster. Let $d(f(y_1), f(y_2)) = js^2$ and $d(y_1, y_2) = JS^2$. The self similar transformation is virtually contraction map, we then have an equation similar to (1)

$$d(f(y_1), f(y_2)) = r \cdot d(y_1, y_2) \qquad , \qquad (6)$$

$$js^2 = rJS^2 \qquad (7)$$

(7) shows that there is a quantitative relationship between $js^2$ and $JS^2$. Note that the outside space of a cluster is its embedded space, which always is the Euclidean space of dimensions $N$. After the self similar transformations and rescaling on the $m$ th hierarchy a cluster which was originally called the ($m$-1)th order cluster has changed into a new lattice, which is called a ($m$-1)th order lattice. As a lattice it lies the inside of the $m$ th order cluster, so its outside space is just the inside space of the $m$ th order cluster of dimensions $D$. For an observed object, whenever it serves as a lattice spin the interacting energy equals $js^2$; as a cluster spin, however, the energy is $JS^2$. As mentioned previously, a lattice in a cluster of dimensions $D$ can be equivalently regarded as a lattice in a super cube of dimensions $D$ with edge $n$. It is well known that a coordination number for a $D$-dimensional cube is $2D$, so that the total magnitudes of interacting energies of a lattice spin with all its nearest neighbors inside the $m$ th order cluster equal $2Djs^2$. As a ($m$-1)th order cluster spin before rescaling, however, the total interacting energies of it with all its nearest neighbors, in its outside space, equal $ZJS^2$, where $Z$ is a coordination number of the cluster spin. In fact, the lattice spin and the cluster spin are the same observed object described by two above artificial versions, hence these different descriptions must be equivalent in magnitudes, which leads to an equality

$$ZJS^2 = 2Djs^2 \quad , \tag{8}$$

where $s^2 = 1$. Comparing (8) with (7), we find

$$r = Z/(2D) \tag{9}$$

As the constant $L$ in (1), $r$ is limited indeed. Let $K = J/(k_B T)$, $k_B$ the Boltzman constant, $T$ temperature, (8) becomes

$$ZKS^2 = 2Djs^2/(k_B T) \tag{10}$$

For a given system, $Z$ is constant, $s^2 = 1$, $j$ is the coupling constant in the space of dimensions $D$, which means $K$ and $S^2$ also relate to $D$, and they are not independent of each other. In renormalization group theory, Wilson preferred to suppose $S^2 = s^2$, namely, the cluster spins always will not change, so that $K$ is only determined by $D$ at the critical temperature due to (10). Since the fractal dimension $D$ is an edge function, $K$ is certainly controlled by the edges. In fact, $K_c$ is merely determined by the edge fixed point. We think that there is no enough evidence to guarantee the supposition of $S^2 = s^2$ to be correct for any systems and clusters no matter what size and shape the cluster will have, and what type of systems such as the plane triangle lattice, the plane square lattice or the cube lattice the cluster lies in. The cluster spins' coupling is virtually some lattice spins' one, which are shown in (7) and (8). Thus we have sufficient reasons to suppose that $J = j$ and $S^2 \neq s^2 = 1$, so (8) becomes

$$ZS^2 = 2D \qquad (11)$$

(11) indicates that different cluster spins with *D* and *Z* have different values. We emphasize here that we derived (8), (10) and (11) by the same observed object, so the *J* and *j* in (8) are only fit for the same system. Each subsystem of a reducible system has its own equalities similar to (8) and (11). The supposition that the cluster spins can change gives us a chance to set up a cluster-spin Gaussian model solved exactly, which will make the calculation of a critical point very simplified and highly accurate.